\documentclass[aps,preprint,nofootinbib]{revtex4}
\usepackage{amsmath}
\usepackage{amssymb}
\usepackage{verbatim}
\begin{document}
\author{R. N. Ghalati}
\email{rnowbakh@uwo.ca}
\affiliation{Department of Applied Mathematics,
University of Western Ontario, London, N6A~5B7 Canada}
\author{D. G. C. McKeon}
\email{dgmckeo2@uwo.ca}
\affiliation{Department of Applied Mathematics,
University of Western Ontario, London, N6A~5B7 Canada}
\title{
A Reexamination of the Canonical Structure of the Einstein-Hilbert Action in First-Order Form}
\date{\today}
\preprint{{\footnotesize UWO\,-TH-\,07/19}}
\begin{abstract}
A canonical analysis of the Einstein-Hilbert action $S_d=\int d^dx \,\sqrt{-g} \,R$\,\,\,\,$(d>2)$ is considered, using the first order 
form with the metric and affine connection as independent fields. We adopt a conservative approach to using the Dirac 
constraint formalism; we do not use equations of motion which are independent of time derivatives and correspond to first class constraints 
to eliminate fields. Applying the 
Dirac procedure, we find that the primary constraints lead to secondary constraints which are equations of motion not involving 
time derivatives, and that those secondary constraints which are first class imply novel tertiary constraints which are also first class.
Once the constraints and their associated gauge conditions are used to
eliminate the non-dynamical degrees of freedom in $S_d$, there are
$d(d-3)$ degrees of freedom left in phase space. We also consider 
the simpler limiting case of the non-interacting graviton in the first order formalism as well as the effect of adding 
the action for a massless scalar field to the Einstein-Hilbert action.
\end{abstract}
\maketitle
\section{Introduction}
Any analysis of the canonical structure of $d$-dimensional Einstein-Hilbert action
\begin{equation}\label{1}
S_d=\int d^dx \sqrt{-g} R
\end{equation}
is greatly complicated by symmetries which appear because 
of the presence of first class constraints. Disentangling the physical degrees of freedom from those that serve 
only to maintain manifest invariance under symmetry transformations is a principal goal of any examination of the 
canonical structure of $S_d$. Having a clear understanding of this structure would be crucial in any quantization 
procedure for the gravitational field.

Einstein's first formulation of general relativity (GR) was solely in terms of the metric $g_{\mu\nu}(x)$, 
but he later \cite {1} showed 
that if $d>2$, then $S_d$ can be considered with the metric and the affine connection $\Gamma^\lambda_{\mu\nu}$ being 
taken as independent. Such a ``first order'' (in derivatives) form of $S_d$ yields the same equations of motion as the 
original ``second order'' form in which $S_d$ depends solely on the metric with the affine connection being identified 
with the Christoffel symbol $\big\{^{\,\,\lambda}_{\mu\nu}\big\}$. (Palatini is often credited with this result \cite {2}.) This is 
because the equation of motion for $\Gamma^\lambda_{\mu\nu}$ when $S_d$ is written in first order form is 
$\Gamma^\lambda_{\mu\nu}=\big\{^{\,\,\lambda}_{\mu\nu}\big\}$ when $d>2$; if $d=2$ then $\Gamma^\lambda_{\mu\nu}$ is 
not uniquely determined by $g_{\mu\nu}$ \cite {3}.

Geometrical variables other than $g_{\mu\nu}$ and $\Gamma^\lambda_{\mu\nu}$ are often used to characterize $S_d$. A second-order 
form can employ the vierbein $e^a_\mu$ while a first order form could use the vierbein and spin connection 
$\omega^\mu_{ab}$. Indeed, if spinors occur in curved space, these geometric quantities must be used \cite {4}. 
It is not even apparent that the formulation of $S_d$ in terms of $g_{\mu\nu}$ and $\Gamma^\lambda_{\mu\nu}$ is 
fully equivalent to that in terms of $e^a_\mu$ and $\omega^\mu_{ab}$ \cite {5}.

The various choices of geometrical quantities to characterize $S_d$ have all been used when analyzing its canonical 
structure. The first order form of $S_d$ in which both $e^a_\mu$ and $\omega^\mu_{ab}$ appear as basic fields has 
been treated \cite {6} using the constraint formalism of Dirac \cite{7,8,9,10,11,12}. If the one basic quantity is 
the spin connection, then the program of ``loop quantum gravity'' can be developed \cite{13,14,15}.

Early treatments of the canonical structure of $S_d$ involve taking the metric or 
the metric and affine connection to be the fundamental fields
\cite{pirani,16,17}. In his analysis of the action in second order
form when $d=4$ \cite{16,17}, Dirac considers the metric to be fundamental and discards those 
portions of $\sqrt{-g}R$ that are the divergence of a vector, keeping only the ``$g\,\Gamma\Gamma$'' part, thereby 
breaking covariance of the Lagrangian. Also, he characterizes each space-like surface in the theory by 
a distinct value of the time parameter $t$.
We adopt the same assumption here, and do not discuss the question of whether in Einstein's theory selecting such a time 
coordinate is feasible.  

The canonical structure of $S_4$ in first order form was first discussed by Arnowitt,
Deser and Misner (ADM) \cite{19,20,21,22}. (See 
also the texts of refs. \cite{23,24}.) In this treatment, all of the equations of motion that do not involve time 
derivatives (the ``algebraic constraints'') are solved for a number of the fundamental fields at the 
level of the Lagrangian. These solutions are 
then used to eliminate these fields from the action, by which one obtains a so called ``reduced'' action; eq.\ (3.3) of 
ref. \cite{25} for example. The canonical analysis of the action
starts at this point \footnote{The first order form of $S_4$, where $g_{\mu\nu}$ and $\Gamma^\lambda_{\mu\nu}$ are the fundamental fields, 
is treated explicitly using this procedure in refs. \cite{20,25}. The approach of ref. \cite{26} to constrained systems 
with first order Lagrangians is much the same as that of
refs. \cite{20,25}.}. Therefore one expects that the four ADM first class
constraints $\mathcal{H}_i$ and $\mathcal{H}$ that are obtained by
working with this form of the Lagrangian lead to generators of a
transformation which is the invariance of the ``reduced''
action, and possibly the gauge invariance of the original EH
action.\footnote{An account of the derivation of the diffeomorphism invariance of
  the EH action in second order form can be found in ref. \cite{33C},
  however, the authors of this paper are unaware of such an account
  for the first order ADM analysis.} 

An essential difference between the canonical analysis of the first
order form of the EH action presented in this paper and that of
previous treatments is that the Dirac constraint formalism is applied 
only using equations of motion corresponding to second class
constraints to eliminate fundamental fields at the Lagrangian
level. As it will be seen, this leads to a constraint structure
sharply distinct from that of ADM. 
As a matter of fact, applying the Dirac constraint analysis to the first order form of $S_2$ has been shown 
\cite{27,28,29,30,31} to lead to a gauge transformation 
that is distinct from a coordinate transformation, even though the Lagrangian is manifestly invariant under a coordinate 
transformation. It might be interesting to make connections between this unexpected result and those of ref. \cite{torre}, where the class 
of all symmetries of the second order Einstein equations of motion in
$d=4$ are studied. It might very well be that having a new symmetry is a feature particular 
to $d=2$.  

In the next section the canonical analysis of $S_d$ in the first order form is given in detail. This program has been outlined 
in ref. \cite{29} although here we use a different set of canonical
variables. The linearized version of 
$S_d$ (i.e. the first order form of the spin-two field \cite{19}) is treated using this formalism in appendix A. 
The effect on the PB algebra of a free massless scalar field is considered in appendix B. The inclusion of a cosmological 
constant, massive scalar fields, Maxwell gauge fields and Yang-Mills fields is considered in \cite{ramin}.
A summary of our results for the constraint structure of the first order EH action appears in ref. \cite{33-khodam}.
  
\section{The EH action in d dimensions} 
In this section we will use the Dirac constraint formalism to analyze the first order form of the EH action
 in $d$ dimensions. Since this is a rather lengthy procedure, subheadings will be used to itemize each of the steps.

\subsection{Choice of Variables}

The EH action of eq. (\ref{1}) when written  in terms of the metric $g_{\mu\nu}$ and the affine connection $\Gamma^\lambda_{\mu\nu}$ is
\begin{equation}\label{2}
S_d=\int d^d x \sqrt{-g} g^{\mu \nu}(\Gamma^\lambda_{\mu \nu , \lambda}-\Gamma^\lambda_{\lambda \mu , \nu} + \Gamma^\lambda_{\mu \nu}
\Gamma^\sigma_{\sigma \lambda}-\Gamma^\lambda_{\sigma \mu} \Gamma^\sigma_{\lambda \nu})\,.
\end{equation}
It is convenient to re express this in terms of the variables 
\begin{eqnarray}\label{3}
h^{\mu \nu} &=& \sqrt{-g} g^{\mu \nu}\\ \label{4}
G^\lambda_{\mu \nu} &=& \Gamma^\lambda_{\mu \nu}-\frac{1}{2}(\delta^\lambda_\nu \Gamma^\sigma_{\mu \sigma}+\delta^\lambda_\mu \Gamma^\sigma_{\nu \sigma})
\end{eqnarray}
so that 
\begin{equation}\label{5}
S_d=\int d^d x \,h^{\mu \nu}\,(G^\lambda_{\mu \nu,\lambda}+\frac{1}{d-1} G^\lambda_{\lambda \mu} 
G^\sigma_{\sigma \nu}-G^\lambda_{\sigma \mu}
G^\sigma_{\lambda \nu})\,.
\end{equation}
If $d \neq 2$, then $g^{\mu\nu}$ can be expressed in terms of $h^{\mu\nu}$ since 
\begin{equation}\label{6}
\det h^{\mu \nu}=-(\sqrt{-g})^{d-2}\,.
\end{equation}
For convenience, we integrate the first term in eq. (\ref{5}) by parts and drop the surface term. 
If $h=h^{00}$, $h^i=h^{0i}$, $\pi=-G^0_{00}$, $\pi_i=-2G^0_{0i}$, $\pi_{ij}=-G^0_{ij}$, $\xi^i=-G^i_{00}$, $\xi^i_j=-2G^i_{j0}$ and 
$\xi^i_{jk}=-G^i_{jk}$, then eq. (\ref{5}) can be written as 
\begin{eqnarray}\label{7}
S_d &=& \int d^d x \bigg[\,\left(\pi h_{,0}+\pi_i h^i_{,0}+\pi_{ij} h^{ij}_{\,\,,0}\right)+\frac{2-d}{d-1}\left(h \pi^2+h^i \pi \pi_i + 
\frac{1}{4} h^{ij} \pi_i \pi_j\right)\\ \nonumber
    &+& \xi^i\left(h_{,i}-h\pi_i- 2 h^j\pi_{ij}\right)\\ \nonumber
    &+& \xi^i_j\,\left(h^j_{,i}+\frac{1}{d-1}\, h\,\pi\, \delta^j_i+
\frac{1}{2(d-1)}\, h^k \,\pi_k \,\delta^j_i -\frac{1}{2} \,h^j \pi_i - h^{jk} \pi_{ik}\right)\\ \nonumber
    &+& \xi^i_{jk} \,\left(h^{jk}_{\,\,\,,i}+\frac{1}{d-1}\, \pi \,(\,\delta^j_i \,h^k
+\delta^k_i\, h^j\,)+\frac{1}{2(d-1)}\,(\delta^j_i \,h^{kl}+\delta^k_i \,h^{jl}\,)\, \pi_l\right) \\ \nonumber
    &+& 
\frac{1}{4} \left(\frac{1}{d-1} \xi^k_k\, \xi^l_{l}-\xi^k_l \,\xi^l_{k}\right)h+
\left(\frac{1}{d-1} \xi^k_{ki}\, \xi^l_{lj}-\xi^k_{li}\,\xi^l_{kj}\right)h^{ij}+ 
\left(\frac{1}{d-1} \xi^k_k\,\xi^l_{li}-\xi^k_l\,\xi^l_{ki}\right)\,h^i \, \bigg]
\end{eqnarray}
At this stage we do not use equations of motion that are independent of time derivatives in order to eliminate any 
of the fields in eq. (\ref{7}), unlike refs. \cite{19,20,21,22,23,24,25}.

We can further simplify the form of eq. (\ref{7}) by first separating the trace of $\xi^i_j$
\begin{equation}\label{8}
\xi^i_j=\bar \xi^i_j+\frac{1}{d-1} \,\delta^i_j\,t
\end{equation}
where $\bar \xi^i_i=0$, and then shifting $\bar \xi^i_j$ to decouple $\bar \xi^i_j$ from $\xi^i_{jk}$ in the action,
\begin{equation}\label{9}
\bar \xi^k_l=\bar \zeta^k_l - \frac{2}{h} \left(\xi^k_{lm}-\frac{1}{d-1}\,\delta^k_l \xi^j_{jm}\right) h^m\,,
\end{equation}
so that eq. (\ref{7}) becomes
\begin{eqnarray}\label{10}
S_d &=& \int d^d x \bigg[\,\left(\pi h_{,0}+\pi_i h^i_{,0}+\pi_{ij} h^{ij}_{\,\,,0}\right)+\frac{2-d}{d-1}\left(h \pi^2+h^i \pi \pi_i + 
\frac{1}{4} h^{ij} \pi_i \pi_j\right)\\ \nonumber
    &+& \xi^i\left(h_{,i}-h\pi_i- 2 h^j\pi_{ij}\right)+\frac{t}{d-1}\left(h^j_{,j}+h\pi-h^{jk}\pi_{jk}\right)\\ \nonumber
    &+& \bar \zeta^i_j\,\left(h^j_{,i} -\frac{1}{2} \,h^j \pi_i - h^{jk} \pi_{ik}\right)-\frac{h}{4} 
        \,\bar \zeta^k_l \,\bar \zeta^l_k\\ \nonumber
    &+& \xi^i_{jk} \,\bigg(h^{jk}_{\,\,\,,i}-\frac{1}{h}(h^j\,h^k)_{,i}+\frac{1}{(d-1)h}\, (\,\delta^j_i \,h^k
        +\delta^k_i\, h^j\,)(h^l_{,l}-\frac{1}{2}h^l\, \pi_l-h^{lm}\,\pi_{lm}+h\,\pi) \\ \nonumber 
    &+& \frac{1}{h} h^j\,h^k\,\pi_i+\frac{1}{h}(h^j\,h^{kl}+h^k\,h^{jl})\,\pi_{il}+\frac{1}{2(d-1)}\,(\delta^j_i \,h^{kl}+
         \delta^k_i \,h^{jl}\,)\, \pi_l\bigg) \\ \nonumber
    &+& H^{ij}\, \left(\xi^k_{li}\,\xi^l_{kj}-\frac{1}{d-1} \,\xi^k_{ki}\, \xi^l_{lj}\right) 
 \bigg]
\end{eqnarray}
where
\begin{equation}\label{11}
H^{ij}=\frac{1}{h}\,h^i\,h^j-h^{ij}\,.
\end{equation}

At this point, it is convenient to replace $h^{ij}$ by $H^{ij}$. If we define
\begin{equation}\label{12}
\omega=\pi-\frac{h^i\,h^j}{h^2}\,\pi_{ij}\,\,\,\,,\,\,\,\,\omega_i=\pi_i+2\,\frac{h^j}{h}\,\pi_{ij}\,\,\,\,,\,\,\,\,\omega_{ij}=-\pi_{ij}
\end{equation}
it follows that
\begin{equation}\label{13}
\pi\, h_{,0}+\pi_i\, h^i_{,0}+\pi_{ij} \,h^{ij}_{\,\,,0}=\omega\, h_{,0}+\omega_i \,h^i_{,0}+\omega_{ij} \,H^{ij}_{\,\,,0}\,.
\end{equation}
The action of eq. (\ref{10}) now becomes 
\begin{eqnarray}\label{14}
S_d &=& \int\,d^d\,x \bigg[\,\omega\, h_{,0}+\omega_i \,h^i_{,0}+\omega_{ij} \,H^{ij}_{\,\,,0}\\ \nonumber
    &+& \,\frac{2-d}{d-1}\bigg(h\,(\omega+\frac{1}{2}\,\frac{h^i\omega_i}{h})^2 - \frac{1}{4} H^{ij}
        (\omega_i+\frac{2\omega_{im}\,h^m}{h})(\omega_j+\frac{2\omega_{jn}\,h^n}{h})\bigg)\\ \nonumber
    &+& \bar \xi^i \,\chi_i+\frac{\bar t}{d-1}\,\chi+\lambda^j_i \, \bar \zeta^i_j+ \sigma^{jk}_i\,\xi^i_{jk}- \frac{h}{4}\,\bar \zeta^i_j\,\bar\zeta^j_i+H^{ij}\, \left(\xi^k_{li}\,\xi^l_{kj}-\frac{1}{d-1} \,\xi^k_{ki}\, \xi^l_{lj}\right)
\bigg]\,,
\end{eqnarray}
where
\begin{eqnarray}\label{15}
\chi&=&h^j_{,j}+h\,\omega-H^{jk}\,\omega_{jk}\,,\\ \label{16}
\chi_i&=&h_{,i}-h\,\omega_i\,,\\ \label{new}
\bar \xi^i&=&\xi^i-\frac{h^jh^k}{h^2}\,\,\xi^i_{jk}\,,\\ \label{new2}
\bar t&=&t+\frac{1}{h}\,\left(\delta^j_ih^k+\delta^k_ih^j\right)\,\xi^i_{jk}\,,
\end{eqnarray}
and
\begin{eqnarray}\label{17}
\lambda^j_i&=&h^j_{,i}-\frac{1}{2}\,h^j\,\omega_i-H^{jk}\,\omega_{ik}\,,\\ \label{18}
\sigma^{jk}_i &=& -H^{jk}_{\,\,,i}+\frac{1}{h}(h^j\,H^{kl}+h^k\,H^{jl})\,\omega_{il}-\frac{1}{d-1}(\delta^j_i\,H^{kl}+
\delta^k_i\,H^{jl})(\frac{1}{2}\,\omega_l+\omega_{lm}\,\frac{h^m}{h})\,.
\end{eqnarray}

At this stage one might decompose $\xi^i_{jk}$ into $\bar \eta^i_{jk}$, $t_i$ and $s^i$ where $ \bar \eta^k_{jk}=0=H^{jk}\bar \eta^i_{jk}$ 
by the equations 
\begin{equation}\label{19}
\xi^i_{jk}=\bar \xi^i_{jk}+\frac{1}{d}\,(\delta^i_j\,t_k+\delta^i_k\,t_j)\,\,\,\,\,\,\,\,\,
\,\,\,\,\,\,\,\,\,\,\,\,\,\,\,\,\,\,\,\,\,\,\,\,\,\,\,\,\,\,\,\,\,\,\,(\bar \xi^k_{jk}=0)
\end{equation}
\begin{equation}\label{20}
\bar \xi^i_{jk}=\bar \eta^i_{jk}+\frac{1}{(d-2)(d+1)}\left(\,d\,H_{jk}\,s^i-(\delta^i_j\,H_{km}+\delta^i_k\,H_{jm})\,s^m\right)
\end{equation}
with $H^{ip}H_{pj}=\delta^i_j$. This however does not simplify the canonical analysis. 

The canonical analysis of the EH action written in the form of eq. (\ref{14}) can now proceed.
\subsection{Primary and Secondary Constraints}
Since eq. (\ref{14}) is first order in the time derivatives, we see immediately that the momenta associated with the fields $\omega$, 
$\omega_i$ and $\omega_{ij}$ are all zero while the momenta associated with $h$, $h^i$ and $H^{ij}$ are $\omega$, $\omega_i$ and
 $\omega_{ij}$ respectively. These constitute a set of $d(d+1)$ primary second class constraints \cite{7,8,9,10,11,12}.

The momenta associated with the fields $\bar t$ and $\bar \xi^i$ also
vanish. As $\bar t$ and $\bar \xi^i$ only enter eq. (\ref{14}) linearly, 
the vanishing of their momenta form a set of $d$ primary first class constraints.

From eq. (\ref{14}) the canonical Hamiltonian is 
\begin{eqnarray}\label{21}
H &=& \,\frac{d-2}{d-1}\bigg(h\,(\omega+\frac{1}{2}\,\frac{h^i\omega_i}{h})^2 - \frac{1}{4} H^{ij}
        (\omega_i+\frac{2\omega_{im}\,h^m}{h})(\omega_j+\frac{2\omega_{jn}\,h^n}{h})\bigg)\\ \nonumber
    &-& \bar \xi^i \,\chi_i-\frac{\bar t}{d-1}\,\chi-
        \lambda^j_i\,\bar \zeta^i_j-\sigma^{jk}_i \xi^i_{jk}+ 
        \frac{h}{4}\,\bar \zeta^i_j\,\bar\zeta^j_i-H^{ij}\, \left(\xi^k_{li}\,\xi^l_{kj}-\frac{1}{d-1} \,\xi^k_{ki}\, \xi^l_{lj}\right)\,.
\end{eqnarray}
In order to describe the dynamics of the gravitational field, instead
of forming the total Hamiltonian by supplementing the
canonical Hamiltonian of eq. (\ref{21}) with primary constraints by
means of Lagrange multipliers, we adopt a different approach. In this
approach, it is not necessary to fix Lagrange multipliers by the emergence of second class
constraints that may arise because of the consistency conditions, but Dirac brackets
are introduced instead of Poisson
brackets and second class constraints are set strongly equal to zero.

Having the momenta associated with $\bar t$ and $\bar \xi^i$ vanish means that these momenta must have a vanishing PB with 
$H$ in eq. (\ref{21}); we thus obtain the secondary constraints 
\begin{eqnarray}\label{22}
\chi &=& 0\\ \label{23}
\chi_i &=& 0\,.
\end{eqnarray}
By using test functions to evaluate the PB of $\chi$ and $\chi_i$ we find that
\begin{equation}\label{24}
\{\chi_i\,,\chi\} = \chi_i
 \end{equation}
while
\begin{equation}\label{25}
\{\chi_i\,,\chi_j\}=0=\{\chi\,,\chi\}\,.
\end{equation}

As has been noted above after eq. (\ref{7}), we do not use equations of motion that have no time derivatives to eliminate fields 
from the action. In particular, two of these equations of motion are the trace of eq.\ (A3) and eq.\ (A4) of ref. \cite{25}, and these are identical 
to our constraints $\chi=\chi_i=0$ of eqs. (\ref{22},\ref{23})\,.

Since by eq. (\ref{25}) it is possible at 
this stage that the constraints $\chi$ and $\chi_i$ are first class, it is necessary to find the PB of these constraints with $H$ to see if there 
are any tertiary constraints. We then must determine if $\chi$ and 
$\chi_i$ continue to be first class once these tertiary constraints are included, and to find what class the tertiary constraints 
belong to. If the tertiary constraints are not seen to be immediately second class, the possibility of ``fourth generation'' 
constraints must be considered and the procedure continues until all constraints are found and classified.

The presence in eq. (\ref{14}) of terms quadratic in $\bar \zeta^i_j$ and $\xi^i_{jk}$ implies that there are also second class 
secondary constraints to be considered. Such constraints do not arise if $d=2$, considerably simplifying the canonical structure 
of the two dimensional EH action \cite{27,28,29,30,31}.
\subsection{Tertiary constraints}
The momenta associated with the traceless quantities $\bar \zeta^i_j$ and the quantity $\xi^i_{jk}$ all vanish; this leads to 
$[(d-1)^2-1]+[\frac{1}{2}d(d-1)^2]=\frac{1}{2}d(d^2-3)$ primary constraints. Taking the PB of these constraints with $H$ 
given in eq. (\ref{21}) results in $\frac{1}{2}d(d^2-3)$ additional secondary constraints, each of which is linear in either 
$\bar \zeta^i_j$ or $\xi^i_{jk}$. Consequently, all of these constraints must be second class; in total there are $d(d^2-3)$ 
second class constraints. The equations of motion that are secondary second class constraints correspond to eq.\ (A2) 
and the traceless part of eq.\ (A3) of ref. \cite{25}.

We can in fact solve these equations of motion and eliminate the variables $\bar \zeta^i_j$ and $\xi^i_{jk}$ in the Hamiltonian 
provided we use the appropriate DB \cite{7,8,9,10,11,12}. Being able to solve these second class constraints in order to eliminate $\bar \zeta^i_j$ and $\xi^i_{jk}$ is quite unlike the 
situation for the first class constraints $\chi$ and $\chi_i$ of eqs. (\ref{22},\ref{23}) which cannot be used to eliminate 
fields in the Dirac constraint formalism.

We first write the portion of the Hamiltonian of eq. (\ref{21}) that generates the secondary second class constraints as
\begin{eqnarray}\label{26}
A &=& -\bar \zeta^i_j\,\lambda^j_i\,+\,\frac{h}{4}\bar \zeta^i_j\,\bar \zeta^j_i\\ \label{27}
B &=& -\xi^i_{jk}\,\sigma^{jk}_i-H^{ij}\, \left(\xi^k_{li}\,\xi^l_{kj}-\frac{1}{d-1} \,\xi^k_{ki}\, \xi^l_{lj}\right)\\ \nonumber
  &\equiv&-\xi^i_{jk}\,\sigma^{jk}_i-\,\xi^k_{lm}\left(M^{\,lm\,\,de}_{\,k\,\,\,\,\,\,c}\right)\xi^c_{de} 
\end{eqnarray}
where
\begin{eqnarray}\label{28}
M^{\,lm\,\,de}_{\,k\,\,\,\,\,\,c} &=& \frac{1}{4}\bigg[H^{me}\left(\delta^l_c\,\delta^d_k-\frac{1}{d-1}\,\delta^l_k\,\delta^d_c\right)
 + H^{md}\left(\delta^l_c\,\delta^e_k-\frac{1}{d-1}\,\delta^l_k\,\delta^e_c\right)\\ \nonumber
 &+& H^{le}\left(\delta^m_c\,\delta^d_k-\frac{1}{d-1}\,\delta^m_k\,\delta^d_c\right)+ H^{ld}\left(\delta^m_c\,\delta^e_k-\frac{1}{d-1}\,\delta^m_k\,\delta^e_c\right) \bigg]\,.
\end{eqnarray}
If
\begin{eqnarray}\label{29}
M^{-1\,x\,\,\,\,k}_{\,\,\,\,\,\,\,\,\,yz\,lm} &=& \frac{1}{2} \bigg[\left(H_{ly}\,\delta^k_z\,\delta^x_m\,+
\,H_{lz}\,\delta^k_y\,\delta^x_m+H_{my}\,\delta^k_z\,\delta^x_l\,+\,H_{mz}\,\delta^k_y\,\delta^x_l \right)\\ \nonumber
 &+& \frac{2}{d-2}(H^{kx}H_{lm}H_{yz})-H^{kx}\,(H_{lz}H_{my}+H_{ly}H_{mz})\bigg]
\end{eqnarray}
then it follows that
\begin{equation}\label{30?}\nonumber
\left(M^{-1\,x\,\,\,\,k}_{\,\,\,\,\,\,\,\,\,yz\,lm}\right)\left(M^{\,lm\,\,de}_{\,k\,\,\,\,\,\,c} \right)=
\frac{1}{2}\delta^x_c\,(\delta^d_y\,\delta^e_z+\delta^d_z\,\delta^e_y)\,.
\end{equation}
The equations of motion for $\bar \zeta^i_j$ and $\xi^i_{jk}$ that follow from $A$ and $B$ in eqs. (\ref{26},\ref{27}) imply that
\begin{eqnarray}\label{30}
\bar \zeta^i_j&=&\frac{2}{h}\,\left(\delta^i_m\delta^n_j-\frac{1}{d-1}\,\delta^i_j\delta^n_m\right)\lambda^m_n\\ \label{31}
\xi^i_{jk}&=&-\frac{1}{2}\,\left(M^{-1\,i\,\,\,\,l}_{\,\,\,\,\,\,\,\,\,jk\,mn}\right) \sigma^{mn}_l\,.
\end{eqnarray}
Substitution of eqs. (\ref{30},\ref{31}) into eqs. (\ref{26},\ref{27}) respectively results in
\begin{equation}\label{32}
A=-\frac{1}{h}\left(\lambda^i_j\,\lambda^j_i-\frac{1}{d-1}\,\lambda^i_i\,\lambda^j_j\right)
\end{equation}
\begin{equation}\label{33}
B=\frac{1}{4}\sigma^{jk}_i\left(M^{-1\,i\,\,\,\,l}_{\,\,\,\,\,\,\,\,\,jk\,mn}\right) \sigma^{mn}_l\,.
\end{equation}
Replacing $A$ and $B$ as given in eqs. (\ref{26},\ref{27}) with $A$ and $B$ as given in eqs. (\ref{32},\ref{33}) 
leads to the Hamiltonian of eq. (\ref{21}) being expressed as a
function that depends exclusively on $(h,\omega)$, $(h^i,\omega_i)$,
$(H^{ij},\omega_{ij})$, $\bar t$ and $\bar \xi^i$. We then drop explicit dependence on $\chi$ and $\chi_i$ occurring in the Hamiltonian of eq. (\ref{21}), leading to the following \emph{weak} Hamiltonian,
\begin{eqnarray}\label{added}
H_{w}&=&h\omega^2+h^i\omega \omega_i-\frac{d-3}{4(d-2)}H^{ij}\omega_i \omega_j-2\frac{h^m}{h}\,H^{ij}\omega_{im}\omega_j-\frac{1}{h}\,H^{ik}H^{jl}\omega_{jk}\omega_{il}\\\nonumber
&+&\frac{1}{h}h^i_{,\,j}h^j\omega_i+\frac{2}{h}\,h^i_{,\,j}H^{jk}\omega_{ik}-\frac{h^i}{h}\,H^{jk}_{\,\,,\,i}\omega_{jk}+\frac{1}{2(d-2)}H_{jk}H^{jk}_{\,\,,\,i}H^{im}\omega_m\\\nonumber
&-&\frac{1}{h}h^i_{,\,j}h^j_{,\,i}+\frac{1}{2}\,H^{jk}_{\,\,,\,i}H_{jq}H^{iq}_{\,,\,k}+\frac{1}{4}\,H^{ip}H_{kr,i}H^{kr}_{\,,\,p}+\frac{1}{4(d-2)}\,H^{ip}H_{jk}H^{jk}_{\,\,,\,i}H_{qr}H^{qr}_{\,\,,\,p}\,.
\end{eqnarray}
Evaluation of the PB of $\chi$ and $\chi_i$ with the Hamiltonian
provides the time change of these constraints \footnote{At this stage, since a set of second class constraints have been set
to zero and solved for a number of fundamental fields in the action,
Poisson Brackets should be replaced by Dirac Brackets. However, as it
is shown in eq. (\ref{43}) below, for the purpose of our calculations we may
safely use PBs instead of DBs.}. However, since we are only interested
in what constraints arise from $\chi$ and $\chi_i$ at this stage, we
may by eqs. (\ref{24},\ref{25}) use $H_w$ instead of the full
Hamiltonian. From $\chi_i$, the following quantity $\bar \tau_i$ is obtained, 
\begin{eqnarray}\label{50}
\bar \tau_i &=& \big\{\chi_i,\int\! dy\, H_w(y)\big\} \\ \nonumber
&=& (H^{pq})_{,i}\,\omega_{pq}-2(H^{pq}\omega_{qi})_{,p}-h^p \omega_{i,p}-\omega_iH^{pq}\omega_{pq}+2H^{pq}\omega_{pq,i}+h^p\omega_{p,i}\,.
\end{eqnarray}
Using the form of $\chi_i$ given in eq. (\ref{16}), we find that this is equivalent to taking
\begin{eqnarray}\label{51}
\tau_i&=&h(\frac{1}{h}H^{pq}\omega_{pq})_{,i}+H^{pq}\omega_{pq,i}-2(H^{pq}\omega_{qi})_{,p}\\ \nonumber
&=&\bar \tau_i+h^p\left[\left(\frac{\chi_p}{h}\right)_{\!,\,i}-\left(\frac{\chi_i}{h}\right)_{\!,\,p}\right]-\frac{\chi_i}{h}\,H^{pq}\omega_{pq}
\end{eqnarray}
\noindent to be the tertiary constraint following from $\chi_i$. 
Similarly, if $\big\{\chi(x),\int dy H_w(y)\big\} \approx 0$ we find that
\begin{equation}\label{52}
\bar \tau=H_w+\partial_i\,\delta^i \approx H+\partial_i\,\delta^i
\end{equation}
must weakly vanish. Remarkably, $\bar \tau$ equals the {\it{weak}} Hamiltonian of eq. (\ref{added}) plus the divergence of a vector
\begin{equation}\label{53}
\delta^i=-H^{ij}_{\,\,,j}+\frac{1}{h}(h^ih^j),_{j}+2h^i\omega-H^{ij}(\omega_j+\frac{2\omega_{jm}h^m}{h})-\frac{d}{d-1}\,\frac{h^i}{h}\,\chi\,.
\end{equation}
Carefully combining terms in the Hamiltonian $H_w$ of eq. (\ref{added}) and $\partial_i\, \delta^i$, it follows that
\begin{eqnarray}\label{54}
\bar \tau &=& \tau+\frac{h^i}{h}\,\tau_i+\frac{h^i}{h}\,\chi_{,i}-\frac{h^jh^i_{,j}}{h^2}\,\chi_i+\frac{H^{jk}\omega_{jk}}{h}\,\chi-\frac{h^i\omega}{h}\,\chi_i\\ \nonumber
&+&\frac{2}{h^2}\,h^kH^{ij}\omega_{ik}\,\chi_j+\omega\chi-\frac{d}{d-1}\,\left(\frac{h^i}{h}\,\chi\right)_{\!\!,\,\,i}\,,
\end{eqnarray}
where
\begin{eqnarray}\label{55}
\tau &=& -H^{ij}_{,ij}-(H^{ij}\omega_j)_{,i}-\frac{d-3}{4(d-2)}\,H^{ij}\omega_i\omega_j+\frac{1}{2(d-2)}\,H_{kl}H^{kl}_{\,\,,i}H^{ij}\omega_j\\ \nonumber
&-&\frac{1}{h}H^{ik}H^{jl}(\,\omega_{jk}\,\omega_{il}-\omega_{ik}\,\omega_{jl})+\frac{1}{2}H^{jk}_{\,\,,i}H_{jl}H^{il}_{,k}
+\frac{1}{4}H^{ij}H_{kl,i}H^{kl}_{\,\,,j}\\ \nonumber
&+& \frac{1}{4(d-2)}H^{ij}H_{kl}H^{kl}_{\,\,,i}H_{mn}H^{mn}_{\,\,,j}\,.
\end{eqnarray}
Once again, we are forced to impose a tertiary constraint in order to ensure that $d\chi/dt \approx 0$; we take 
this tertiary constraint to be $\tau$ in eq. (\ref{55}).

An alternate way of obtaining the tertiary constraints is to work with the Hamiltonian in the form of eq. (\ref{21}) without
 eliminating $\bar \zeta^i_j$ and $\xi^i_{jk}$. This means using a DB in place of a PB if $\bar \zeta^i_j$ or
 $\xi^i_{jk}$ are involved.

To illustrate how this works, it is convenient to consider a simplified model in which we have the action 
\begin{equation}\label{34}
S=\int dt \{\,p_i\dot q_i-[\,H_0(q_i,p_i)+\lambda_A\,\chi_A(q_i,p_i)+f^a_I(q_i,p_i)
\,Q^a_{I}-\frac{1}{2}Q^a_I\,\,g^{ab}_{IJ}(q_i)\,\,Q^b_J\,]\,\}\,,
\end{equation}
where
\begin{equation}\label{35}
\{\chi_A,\chi_B\} = C_{ABC}\,\chi_C\,.
\end{equation}
Eqs. (\ref{34},\ref{35}) are analogues of eqs. (\ref{14},\ref{24}-\ref{25}) respectively, with $q_i$ representing $(h,h^i,H^{ij})$, $p_i$ 
representing $(\omega,\omega_i,\omega_{ij})$, $Q^a_I$ representing $(\bar \zeta^i_j$, $\xi^i_{jk})$ and $\lambda_A$ representing 
$(\bar t,\bar \xi^i)$. The momenta conjugate to $Q^a_I$ and $\lambda_A$ ($P^a_I$ and $\pi_A$) are zero; these primary constraints immediately 
give rise to the secondary constraints
\begin{equation}\label{36}
\bar \theta^a_I \equiv f^a_I(q_i,p_i)-g^{ab}_{IJ}(q_i)\,Q^b_J=0
\end{equation}
and
\begin{equation}\label{37}
\gamma_A = \chi_A(q_i,p_i)=0\,.
\end{equation}
The constraints $\theta^a_I=P^a_I=0$ and $\bar \theta^a_I$ of eq. (\ref{36}) are obviously second class while $\gamma_A$ of eq. (\ref{37}) 
may be first class on account of eq. (\ref{35}). (Subsequent tertiary constraints may change these constraints to second class.)

In order to eliminate the second class constraints from the action, we need to form the appropriate DBs. 
Since
\begin{equation}\label{38}
\{\theta^a_I,\bar \theta^b_J\}=g^{ab}_{IJ}\,\delta_{IJ}
\end{equation}
and 
\begin{eqnarray}\label{39}
\{\bar \theta^a_I,\bar \theta^b_J\} &=& \{f^a_I(q_i,p_i)-g^{am}_{IK}(q_i)\,Q^m_K\,,\,f^b_J(q_i,p_i)-g^{bn}_{JL}(q_i)Q^n_L\}\\ \nonumber
&\equiv& M^{ab}_{IJ}
\end{eqnarray}
then the matrix $d_{\alpha \beta}=\{\chi_\alpha,\chi_\beta\}$, where $\chi_\alpha$ and $\chi_\beta$ are second class constraints to be eliminated \cite{7,8,9,10}, takes the form
\begin{equation}\label{40}
d=\begin{pmatrix}
0 & g_1 & 0 & 0\\
-g_1 & M_{11} & 0 & M_{12}\\
0 & 0 & 0 & g_2\\
0 & -M_{12}  & -g_2  & M_{22} 
\end{pmatrix}  
\end{equation} 
with the indices $I$ and $J$ in eq. (\ref{36}) taking on two values,
corresponding to $\bar \zeta^i_j$ and $\xi^i_{jk}$. Using the relation \cite{9} 
\begin{multline}\label{41}
{\begin{pmatrix} A & B \\ C & D \end{pmatrix}}^{-1}
=
{\bigg[\begin{pmatrix}I & B \\ 0 & D \end{pmatrix} \begin{pmatrix} A-BD^{-1}C & 0 \\ D^{-1}C & I \end{pmatrix}\bigg]}^{-1}
\\=
\begin{pmatrix} (A-BD^{-1}C)^{-1} & -(A-BD^{-1}C)^{-1}BD^{-1} \\
-D^{-1}C(A-BD^{-1}C)^{-1} & D^{-1}C(A-BD^{-1}C)^{-1}BD^{-1}+D^{-1} \end{pmatrix}
\end{multline}
we find that
\begin{equation}\label{42}
d^{-1}=\begin{pmatrix}
g_1^{-1}M_{11}\,g_1^{-1} & - g_1^{-1} & g_1^{-1}M_{12}\,g_2^{-1}  & 0\\
g_1^{-1} & 0 & 0 & 0\\
-g_2^{-1}M_{12}\,g_1^{-1} & 0 & g_2^{-1}M_{22}\,g_2^{-1} & -g_2^{-1}\\
0 & 0  & g_2^{-1}  & 0 
\end{pmatrix}\,.
\end{equation}
From eq. (\ref{42}), the definition of the DB \cite{7,8,9,10}, 
\begin{equation*}
\label{A14}
\left\{A,B\right\}^*=\left\{A,B\right\}-\left\{A,\chi_\alpha\right\}\,\,\,\left(d^{-1}\right)^{\alpha\beta}\left\{\chi_\beta,B\right\}\,,
\end{equation*}
shows that in this system
\begin{equation}\label{43}
\{q_i,p_j\}^*=\delta_{ij}\,,
\end{equation}
\begin{equation}\label{44}
\{q_i,Q_{Ia}\}^*=\{q_i,\bar \theta^c_I\}\,(g^{-1})^{ca}_{IJ}\,\delta_{IJ}\,,
\end{equation}
\begin{equation}\label{45}
\{p_i,Q_{Ia}\}^*=\{p_i,\bar \theta^c_I\}\,(g^{-1})^{ca}_{IJ}\,\delta_{IJ}\,,
\end{equation}
\begin{equation}\label{46}
\{Q^a_I,Q^b_J\}^*=(g^{-1})^{am}_{IK}\,\,M^{mn}_{KL}\,\,(g^{-1})^{nb}_{LJ}\,.
\end{equation}
An explicit calculation shows that the matrices $M^{mn}_{KL}$ in eq. (\ref{46}) are non local. 
This makes the use of eq. (\ref{46}) somewhat ambiguous, but we will see that in the process of evaluating 
the tertiary constraints corresponding to the secondary constraints $\chi$ and $\chi_i$ we luckily don't need them. In fact, using the constraint $\bar \theta^a_I$ to express the Hamiltonian that follows from eq. (\ref{34}) in the form 
\begin{equation}\label{47}
H=H_0+\lambda_A\,\chi_A+\frac{1}{2}\,Q^a_I\,g^{ab}_{IJ}\,Q^b_J\,,
\end{equation}
it follows from eqs. (\ref{35},\ref{44},\ref{45}) that
\begin{equation}\label{48}
\frac{d\chi_A}{dt} \approx \{\chi_A,H\}^* \approx \{\chi_A,H_0\}+\{\chi_A,\bar \theta^a_I\}\,Q^a_I+\frac{1}{2}\,Q^a_I\,\{\chi_A,g^{ab}_{IJ}\}\,Q^b_J\,. 
\end{equation}
Eq. (\ref{48}) can be used to find the tertiary constraints $\tau_i$ and $\tau$ that follow from the secondary constraints of eqs. (\ref{22},\ref{23}).

It is now necessary to see how the constraints $\chi$, $\chi_i$, $\tau$ and $\tau_i$ are to be classified, and if any 
further ``fourth generation'' constraints are required in order to ensure that $\tau$ and $\tau_i$ have weakly vanishing time derivatives.
\subsection{Algebra of Constraints}
In addition to the PB of eqs. (\ref{24},\ref{25}), one can show easily that
\begin{equation}\label{56}
\{\chi,\tau_i\}=0\,.
\end{equation}
Another direct calculation (one that is somewhat more difficult) leads to
\begin{equation}\label{57}
\{\chi,\tau\}=\tau\,.
\end{equation}
It is also possible to show that
\begin{equation}\label{58}
\{\chi_i,\tau\}=0\,
\end{equation}
and
\begin{equation}\label{59}
\{\chi_i,\tau_j\}=0\,.
\end{equation}

A rather involved calculation leads to
\begin{equation}\label{60}
f\{\tau_i,\tau_j\}g=g(\partial_jf)\tau_i-f(\partial_ig)\tau_j\,,
\end{equation}
where $f$ and $g$ are test functions \cite{GS}. More explicitly, eq. (\ref{60}) can be written as
\begin{eqnarray}\label{61}
\int dx\,dy\,&f(x)& \{\tau_i(x),\tau_j(y)\}\,\,g(y) \\ \nonumber
&=& \int dx \left[g(x)\left(\partial_j f(x)\right)\tau_i(x)-f(x)
\left(\partial_i g(x)\right)\tau_j(x)\right]\\ \nonumber
 &=& \int dx\,dy \left[f(x)\left(-\partial^x_j\,\delta(x-y)\tau_i(y)+\tau_j(x)\,\partial^y_i\,\delta(x-y)\right)g(y)\right]\,,
\end{eqnarray}
so that we have the non-local PB
\begin{eqnarray}\label{62}
\{\tau_i(x),\tau_j(y)\}=-\partial^x_j\,\delta(x-y)\tau_i(y)+\tau_j(x)\,\partial^y_i\,\delta(x-y)\,.
\end{eqnarray}
This is identical to the PB of the constraints $\mathcal{H}_i$ appearing in refs. \cite{19,20,21,22,25}, even though $\tau_i$ and $\mathcal{H}_i$ are distinct.

As mentioned, a disadvantage of the Dirac Brackets introduced in Section C is that the matrices $M^{mn}_{KL}$ occurring in eq. (\ref{46}) are non local. 
Therefore, at the stage developed in this paper, it is not
straight forward how the PBs of the tertiary 
constraints $\tau_i$ and $\tau$, and of $\tau$ and $\tau$, and their
time derivatives must be computed using them. As a result, in order to
find these 
PBs of first class constraints and their 
time derivatives, we use the alternative method where we solved $\xi^i_{jk}$ and $\bar \zeta^i_j$ in terms of $h$, $h^i$, $H^{ij}$, $\omega$, $\omega_i$ and 
$\omega_{ij}$ by means of the second class constraints occurring in the theory. 

When computing the PBs $\{\tau,\tau\}$ and $\{\tau_i,\tau\}$ we are confronted with huge expressions which are rather difficult to arrange into combinations of 
first class constraints. However, it is indeed necessary to show that these PBs are weakly zero if $\tau$ and $\tau_i$ are to be identified as first class constraints. 
It must also be shown that the time derivatives of these tertiary constraints do not lead to fourth generation constraints. We now explain how these two problems are 
intimately connected, and how this connection helps to resolve the algebraic difficulty of computing the PBs $\{\tau,\tau\}$ and $\{\tau_i,\tau\}$.

The observation that the first class constraint $\tau$ of eq. (\ref{55}) weakly differs from the Hamiltonian by a total divergence $\partial_i \delta^i$ is useful. 
Based on the number of degrees of freedom in the non interacting graviton field, one expects that all tertiary constraints are first class and therefore no higher 
generation of constraints should arise. One then concludes that the
time change of $\tau$ and $\tau_i$, and therefore,  $f\big\{\tau_i,\int H_w dy \big\}$ and 
$f\big\{\tau,\int H_w dy \big\}$, where $H_w$ is given by eq. (\ref{added}) should be written as a linear combination of first class constraints. But since $\tau \approx H_w+\partial_i \delta^i$, one concludes that 
$f\big\{\tau_i,\int \tau dy \big\} \approx f\big\{\tau_i,\int H_w dy \big\}$ and also that $f\big\{\tau,\int \tau dy \big\} \approx f\big\{\tau,\int H_w dy \big\}$.  
In other words, $f\big\{\tau_i,\int \tau dy \big\}$ and $f\big\{\tau,\int \tau dy \big\}$ should be expressible in terms of first class constraints. These expressions, 
though still enormous, have turned out to be manageable. They not only lead us to first class expressions for the time change of $\tau$ and $\tau_i$, but also infer how 
some of the terms appearing in $f\big\{\tau_i,\tau\big\}g$ and $f\big\{\tau,\tau\big\}g$ can be written in terms of linear combinations of constraints.

Having these considerations in mind, we first compute the time change of the constraint $\tau$ and find that it is given by a linear combination of constraints  
\begin{equation}\label{66-aaa}
f\big\{\tau,\int dy\, H_w\big\}=\partial_if\frac{H^{ij}}{h^2}\left(h\tau_j-H^{mn}\omega_{mn}\chi_j+2H^{mn}\omega_{mj}\chi_n\right)\,.
\end{equation}
The structure of the last two expressions on the right hand side of this equation resembles that of the last two terms in the constraint $\tau_i$ of eq. (\ref{51}). 
This suggests a redefinition of the constraint $\tau_i$ in order to
obtain a simpler algebra that might be closer to that of the ADM
algebra\footnote{
This is
\begin{eqnarray*}
\big \{\mathcal{H}(x),\mathcal{H}(y)\big \}&=&\big (\mathcal{H}_i(x)+\mathcal{H}_i
(y)\,\big)\,\partial^x_i\delta(x-y)\,,\\
\big \{\mathcal{H}_i(x),\mathcal{H}(y)\big \}&=&
\mathcal{H}(x)\,\partial^x_i\,\delta(x-y)\,,\\
\big \{\mathcal{H}_i(x),\mathcal{H}_j(y)\big \}&=& \left(\mathcal{H}_i(y)\partial^x_j+\mathcal{H}_j(x)\partial^x_i\right)\,\delta(x-y)\,.
\end{eqnarray*}
}, but so far this 
effort has not been successful.
Using eq. (\ref{66-aaa}) for the time change of $\tau$, we are aided in finding that the PB  $\{\tau,\tau\}$ is   
\begin{equation}\label{66}
f\big\{\tau,\tau\big\}g=\left(g\partial_if-f\partial_ig\right)\frac{H^{ij}}{h^2}\left(h\tau_j-H^{mn}\omega_{mn}\chi_j+2H^{mn}\omega_{mj}\chi_n\right)\,.
\end{equation}
In much the same way, the time change of $\tau_i$ is expressible as a linear combination of constraints,
\begin{eqnarray}\label{67aaa}
f\big\{\tau_i,\int dy \,H_w\big\} &=&  \frac{(fh)_{,i}}{h}\,\tau+\frac{d-3}{2(d-2)}\,f\left(\frac{1}{h}\,H^{kl}\omega_l\chi_i\right)_{,\,k}
-\frac{d-3}{2(d-2)}fH^{kl}\omega_l\left(\frac{\chi_k}{h}\right)_{,\,i}\\ \nonumber 
&-&\frac{1}{2(d-2)}\,f\left(H^{mj}H_{kl}H^{kl}_{,j}\frac{\chi_i}{h}\right)_{,m}+\frac{1}{2(d-2)}\,fH^{ml}H_{jk}H^{jk}_{,l}\left(\frac{\chi_m}{h}\right)_{,i}
\end{eqnarray}
and this helps us show that
\begin{eqnarray}\label{67}
f\big\{\tau_i,\tau\big\}g &=& g \frac{(fh)_{,i}}{h}\,\tau-fg_{,i}\tau-\frac{d-3}{2(d-2)}fgH^{kl}
\omega_k\left(\frac{\chi_l}{h}\right)_{,i}\\ \nonumber
&-&\frac{d-3}{2(d-2)}gf_{,k}H^{kl}\,\omega_l\left(\frac{\chi_i}{h}\right)+f_{,k}\,g_{,l}\,H^{kl}\left(\frac{\chi_i}{h}\right)\\ \nonumber
&+& f g_{,k} H^{kl}\left(\frac{\chi_l}{h}\right)_{,i}+\frac{1}{2(d-2)}g\,f_{,m}\,H^{mn}H_{kl}H^{kl}_{\,\,,n}\left(\frac{\chi_i}{h}\right)\\ \nonumber
&+& \frac{1}{2(d-2)}fgH^{mn}H_{kl}H^{kl}_{\,\,,m}\left(\frac{\chi_n}{h}\right)_{,i}\,.
\end{eqnarray}
Eqs. (\ref{24},\ref{25},\ref{56}-\ref{59},\ref{62},\ref{66},\ref{67}) all show that amongst themselves, 
$\chi$, $\chi_i$, $\tau$ and $\tau_i$ are first class and their PB
algebra is highly unusual. 

The Jacobi identities for the PBs of the first class constraint triplets $\left(\tau,\tau,\chi\right)$, $\left(\tau,\tau_i,\chi\right)$ and $\left(\tau,\tau,\tau\right)$ 
have been verified by explicit computation, providing a non-trivial consistency test for the PBs of eqs. (\ref{66}) and (\ref{67}).
\section{Discussion}
We have found the complete constraint structure for the action $S_d$ of eq. (\ref{5}) if $d>2$. In particular, we have the 
$d(d+1)$ primary second class constraints resulting from the identification of $-G^0_{00}$, $-2G^0_{0i}$ and $-G^0_{ij}$ with 
the canonical momenta conjugate to $h$, $h^i$ and $h^{ij}$. We have already noted that there are $d(d^2-3)/2$ primary second 
class constraints associated with the vanishing of the canonical momenta for $\bar \zeta^i_j$ and $\xi^i_{jk}$ and that these 
in turn lead to a further $d(d^2-3)/2$ secondary second class constraints associated with the equations of motion for $\bar \zeta^i_j$ 
and $\xi^i_{jk}$. In total there then are $d(d+1)+d(d^2-3)=d^3+d^2-2d$ second class constraints. We also have $d$ primary first class 
constraints (the momenta associated with $\bar t$ and $\bar \xi^i$) as well as $d$ secondary first class constraints ($\chi$ and $\chi_i$) and $d$
tertiary first class constraints ($\tau$ and $\tau_i$). When we include the gauge conditions associated with each of these $3d$ 
first class constraints, there are $3d+3d+d^3+d^2-2d=d(d^2+d+4)$ restrictions on the $d(d+1)^2$ variables in phase space 
(the $h^{\mu\nu}$, $G^\lambda_{\mu\nu}$ and their conjugate momenta). There are thus $d(d+1)^2-d(d^2+d+4)=d(d-3)$ independent 
degrees of freedom in phase space. If $d=3$, there are no degrees of freedom while if $d=4$, there are the two polarizations 
of the graviton as well as their conjugate momenta. This is in agreement with the expectations of ref. \cite{29}.

In the ADM approach to the first order action of eq. (\ref{2}) (refs. \cite{20,25}) in $d=4$ dimensions, six of the ten 
components of the metric fields are dynamical and the remaining four become Lagrange multipliers, related to the ``lapse'' 
and ``shift'' functions. Thirty equations of motion that correspond to the secondary constraints of eqs. (\ref{22},\ref{23},
\ref{30},\ref{31}) do not contain time derivatives and are used to
eliminate components of the affine connections. (The first four of these
equations, $\chi=\chi_i=0$, which are first class constraints in our
treatment, if used to eliminate $\omega$ and $\omega_i$ would reduce
the Hamiltonian of eq. (\ref{added}) to the ADM Hamiltonian, eq.\ (3.3)
of ref. \cite{25}.) Furthermore, once the elimination has taken place, 
all of the affine connections $\Gamma^\mu_{00}$ disappear from the
action and are not considered to be dynamical in the ADM approach. (In
the analysis presented in this paper, $\Gamma^i_{00}$  and
$\Gamma^0_{00}$ are associated with the Lagrange multipliers $\xi^i$
and $t$ respectively.) There are then six remaining components of the 
affine connection that form the momenta conjugate to those components of the metric which are dynamical. When these constraints are combined with their
associated gauge conditions, only the two transverse degrees of
freedom associated with the metric plus their conjugate momenta remain
in phase space. We thus see how the analysis presented in this paper, which uses exclusively the Dirac constraints formalism \cite{7,8,9,10,11,12}, is 
related to the more conventional ADM approach to the canonical structure of $S_d$ of eq. (\ref{1}) \cite{19,20,21,22,23,24,25}\,. 

The relationship between the Dirac approach and that of ref. \cite{26} is discussed in ref. \cite{31A}. There it is 
shown how the Dirac procedure can be cast into a form that is the same as that of ref. \cite{26}. However, there does not make it clear how to classify the constraints 
that arise at each step of ref. \cite{26}, or if the PB algebra of the resulting constraints is identical to that of the constraints obtained by applying the Dirac procedure exclusively.
Consequently, it is important to know the connection between the ADM
constraints and the constraints found in this paper. In attempting to
do this, we might try to find linear combinations of constraints that simplify our algebra.
As a matter of fact, by replacing $\bar \tau$ in eq. (\ref{54}) by $\tau$ 
in eq. (\ref{55}), the
algebra of PB of constraints has already been simplified, as the PB $\big\{\chi_i,\bar \tau\big\}$ is non local and $d$-dependent,
\begin{equation}\label{additional}
f\big\{\chi_i,\bar \tau\big\}g=fg\,\tau_i-\frac{2}{d-1}\,f\partial_ig\,\chi-\frac{h^j}{h}\,f\partial_ig\,\chi_j-\frac{h^j}{h}f\partial_jg\,\chi_i\,, 
\end{equation}
in contrast to eq. (\ref{58}). It is quite possible that even more simplification occurs if the first class constraints were combined 
in a judicious manner. For example, if $\tilde \chi_i=\chi_i/h$ then $\big\{\tilde \chi_i,\chi\big\}=0$ in place of eq. (\ref{24}). 
It remains to be seen if the PBs of eqs. (\ref{62}-\ref{67}) could be similarly simplified \footnote{A way of simplifying the ADM PB algebra 
is given in refs. \cite{33-2,34-2}.}. This could also possibly provide some 
insight into the geometrical significance of the first class constraints $\chi$, $\chi_i$, $\tau$ and 
$\tau_i$ which is not immediately apparent. We note though that no
matter what the most convenient form of the first class constraints
may be, there will always be tertiary constraints which will
necessarily lead to transformations involving second derivatives of
the gauge functions. This is to be expected as the coordinate
transformation of the affine connection lead to such second
derivatives. If the second order form of the EH action were
considered, then only secondary first class constraints would arise as
in ref. \cite{SN}. In the first order formalism in which the vierbein
and $e^\mu_a$ and the spin connection $\omega^\mu_{ab}$ are the
independent fields, only secondary constraints should arise, as both
the vierbein and affine connection are covariant under a coordinate
transformation, and hence only first derivatives of the gauge
functions occur, consistent with the results of ref. \cite{6}.

The most obvious problem that follows from our analysis that should be addressed is the question of finding the gauge transformation associated with the first class constraints.
Having the gauge 
invariance for the fields $h^{\mu\nu}$  and $G^\lambda_{\mu\nu}$ makes
it possible to apply the quantization procedure outlined in refs. \cite{34,35,36}. When this was done in two dimensions \cite{37}, the transformations to be considered were other than diffeomorphism and the resulting radiative effects appear to cancel. It would be quite interesting to see what radiative effects 
follow from eq. (\ref{5}), especially since it is only a cubic polynomial in the fields.

Extending our analysis to systems which include Bosonic matter fields
such as massive scalar fields, Maxwell gauge fields and Yang-Mills
fields has been done in \cite {ramin}. Having a coupling between the
gravitational field and spinors would mean \cite{4} that the canonical
analysis would have to be done using the vierbein and spin connection
as geometrical fields as in ref. \cite{6}. This analysis would be quite
distinct from the one done here in terms of the metric and affine
connection.

\section{Acknowledgments}
We would especially like to thank N. Kiriushcheva and S.V. Kuzmin for extensive discussions on many aspects of this work. R. N. Ghalati would also like to thank K. Kargar for discussions. D.G.C. McKeon is grateful to F.T. Brandt and 
T.N. Sherry for their help with portions of the analysis. R. Macleod had a useful suggestion.

\appendix
\newpage
\section{Canonical Analysis of the spin-two field in first order formalism}
In this appendix we examine the canonical structure of linearized gravity in first order form using the Dirac constraint 
formalism. It differs in interesting ways from the structure of the full theory outlined in the body of this paper. 
Various aspects of this problem are considered in refs. \cite{41,42,43,44,45,46,47,48,49}.

In order to linearize the action of eq. (\ref{5}), we merely replace it by \cite{19}
\begin{equation}\label{B1}
\tilde S_d=\int d^d x \left[\,h^{\mu \nu}\,G^\lambda_{\mu \nu,\lambda}+\eta^{\mu\nu}\left(\frac{1}{d-1} G^\lambda_{\lambda \mu} 
G^\sigma_{\sigma \nu}-G^\lambda_{\sigma \mu}G^\sigma_{\lambda \nu}\right)\right]
\end{equation}
where $\eta^{\mu\nu}=diag(-,+,+,\ldots,+)$ is the flat space metric.

Eqs. (\ref{28},\ref{29}) can be used to solve the equations of motion of $G^\lambda_{\mu\nu}$, expressing $G^\lambda_{\mu\nu}$ in 
terms of $h_{\mu\nu}$. Using this in order to eliminate $G^\lambda_{\mu\nu}$ in eq. (\ref{B1}), we find that
\begin{equation}\label{B2}
\tilde S_d=\frac{1}{2}\int d^dx\left[h^{\mu\lambda}_{\,\,\,,\sigma}\,h_{\mu\,\,,\lambda}^{\,\,\,\sigma}-
\frac{1}{2} h^{\mu\nu}_{\,\,\,,\lambda}h_{\mu\nu,}^{\,\,\,\,\,\,\lambda}+\frac{1}{2(d-2)}
h^\mu_{\,\,\mu,\lambda}h^{\nu\,\,\lambda}_{\,\,\nu,}\right]
\end{equation}
provided $d \neq 2$. (This case will be dealt with presently.) If $d=4$, eq. (\ref{B2}) is seen to be the action for a spin-two 
field appearing in refs. \cite{50,51}.

The momentum conjugate to $h^{00}=h$, $h^{0i}=h^i$ and $h^{ij}$ (upon integration by parts in the first term of eq. (\ref{B1})) are 
respectively 
$$\label{B3-B5}
\pi=-G^0_{00}\,\,\,\,\,\,,\,\,\,\,\,\,\pi_i=-2G^0_{0i}\,\,\,\,\,\,,\,\,\,\,\,\,\pi_{ij}=-G^0_{ij}\,.\eqno(B3-B5)
$$
If now we define
$$\label{B6-B8}
\xi^k=G^k_{00}\,\,\,\,,\,\,\,\,\,\xi^i_j=2G^i_{j0}=\bar \zeta^i_j+\frac{1}{d-1}\,\delta^i_j\,t\,\,\,\,\,,\,\,\,\,\,
\xi^i_{jk}=G^i_{jk}\,\eqno(B6-B8) 
$$
\addtocounter{equation}{6}
where $t=\xi^i_i$, then the canonical Hamiltonian is
\begin{eqnarray}\label{B9}
H &=& \pi h_{,0}+\pi_ih^i_{\,,0}+\pi_{ij}h^{ij}_{\,\,,0}-L\\ \nonumber
&=&\,\frac{2-d}{d-1}\left(\pi^2-\frac{1}{4}\pi_i\pi_l\right)+\xi^k(\pi_k+h_{,k})+\frac{t}{d-1}(-\pi_{ii}-\pi+h^i_{,i})\\ \nonumber
&+& \left[\bar \zeta^i_j(-\pi_{ij}+h^j_{,i})-\frac{1}{4}\bar \zeta^i_j\,\bar \zeta^j_i\right]+\left[\xi^i_{jk}(h^{jk}_{\,\,,i}+\frac{1}{d-1}\,\delta^j_i\pi_k)+\xi^i_{jk}\,\xi^j_{ik}-\frac{1}{d-1}\xi^i_{ik}\,\xi^j_{jk}\right]\,.
\end{eqnarray}
Many features of the Hamiltonian of eq. (\ref{B9}) resemble those of eq. (\ref{21}). In particular, the momenta
associated with $t$ and $\xi^i$ vanish; these primary first class constraints result in the secondary constraints
\begin{eqnarray}\label{B10}
\chi_k &=& h_{,k}+\pi_k\\ \label{B11}
\chi &=& h^k_{,k}-\pi-\pi_{kk}\,.
\end{eqnarray}
They have the PB
\begin{equation}\label{B12}
\big\{\chi,\chi\big\}=\big\{\chi_i\,,\chi_j\big\}=\big\{\chi,\chi_j\big\}=0\,,
\end{equation}
in contrast to those of eqs. (\ref{24},\ref{25}). Furthermore, the momenta conjugate to $\bar \zeta^i_j$ and $\xi^i_{jk}$ also vanish. 
These primary constraints are second class as they lead to second class secondary constraints, which are the equations of motion for 
$\bar \zeta^i_j$ and $\xi^i_{jk}$ and these variables enter the equations of motion linearly. Eliminating $\bar \zeta^i_j$ and $\xi^i_{jk}$
from the Hamiltonian of Eq. (B.9) using their equations of motion results in 
\begin{eqnarray}\label{B13}
H&=&\frac{2-d}{d-1}\pi^2+\frac{d-3}{4(d-2)}\pi_i\pi_j+\xi^k(\pi_k+h_{,k})-\frac{t}{d-1}(\pi_{ii}+\pi-h^i_{,i})\\ \nonumber
&+& \left(\pi_{ij}\pi_{ij}-\frac{1}{d-1}\pi_{ii}\pi_{jj}-2\pi_{ij}h^i_{,j}+\frac{2}{d-1}\pi_{kk}h^l_{,l}+\,\frac{d-2}{d-1}h^k_{,k}h^l_{,l}\right)\\ \nonumber
&-& \left(\frac{1}{2(d-2)}h^{ii}_{\,\,,j}\pi_j+\frac{1}{2}h^{jk}_{\,\,,i}h^{ik}_{\,\,,j}+\frac{1}{4(d-2)}h^{mm}_{\,\,,j}h^{nn}_{\,\,,j}
-\frac{1}{4}h^{mn}_{\,\,,j}h^{mn}_{\,\,,j}\right)\,\,.
\end{eqnarray}
One must now see if the secondary constraints of eqs. (\ref{B10}-\ref{B11}) imply any further constraints. As 
\begin{eqnarray}\label{B14}
\big\{H,\chi\big\}&=&\tau \\ \label{B15}
\big\{H,\chi_k\big\}&=&2\left(\frac{d-2}{d-1}\,\chi_{,k}-\tau_k\right)
\end{eqnarray}
there are $d$ tertiary constraints
\begin{eqnarray}\label{B16}
\tau&=&h_{ij\,,\,ij}+\pi_{i,i}\\ \label{B17}
\tau_k &=& \pi_{ii,\,k}-\pi_{ik,i}
\end{eqnarray}
Any pair of the constraints of eqs. (\ref{B10}, \ref{B11}, \ref{B16}, \ref{B17}) have vanishing PB and consequently all are first class. 
There are no fourth generation constraints as
\begin{eqnarray}\label{B18}
\big\{\tau,H\big\}&=&0\\ \label{B19}
\big\{\tau_k,H\big\}&=&-\frac{1}{2}\tau_{,\,k}\,.
\end{eqnarray}

It is now possible to find the gauge transformations implied by the constraints $\chi$, $\chi_k$, $\tau$, and $\tau_k$ as well as 
the first class constraints $\Pi$ and $\Pi_k$, the momenta associated with $t$ and $\xi^k$. The algebra of constraints for this 
spin-two theory is quite simple in comparison to that of the full theory of general relativity, making application of refs. \cite{9,12} 
relatively easy.  For this, we need eqs. (\ref{B12}, \ref{B14}, \ref{B15}, \ref{B18}, \ref{B19}) as well as 
$$\label{B20-B21}
\big\{\Pi,H\big\}=-\chi\,\,\,\,\,,\,\,\,\,\,\big\{\Pi_k,H\big\}=\frac{1}{d-1}\,\chi_k\,. \eqno(B20-B21)
$$
\addtocounter{equation}{2}
This constraint structure is unusual in that derivatives of constraints appear in the PB algebra. A general analysis of the gauge 
transformations implied by the first class constraints in such cases appears in ref. \cite{48}.

The form of the generator of gauge transformations is given by 
\begin{equation}\label{B22}
G=\bar \mu \Pi+\bar \mu^k \Pi_k+\mu \chi+\mu^k\chi_k+\underline \mu \tau+\underline \mu^k\tau_k\,.
\end{equation}
Upon using the formulation of refs. \cite{9,12,48} we find that this
generator leaves the action of eq. (\ref{B1}) invariant provided
\begin{eqnarray}\label{B25}
\dot \mu+\frac{1}{d-1} \bar \mu+2\,\left(\frac{d-2}{d-1}\right)\,\mu_{k,k} &=& 0\\ \label{B26}
\dot \mu^k-\bar \mu^k &=&0\\ \label{B27}
\dot {\underline \mu}-\mu+\frac{1}{2}\underline \mu_{k,k} &=&0\\ \label{B28}
\dot {\underline \mu}_k+2\mu_k &=& 0\,,
\end{eqnarray}
so that $G$ in eq. (\ref{B22}) becomes
\begin{eqnarray}\label{B33}
G &=& \big[-(d-1)\ddot {\underline \mu}+\frac{1}{2}\,(d-3)\dot{\underline \mu}_{\,k,k}\big]\Pi+\big[-\frac{1}{2}\,\ddot{\underline \mu}^k\big]\,\Pi_k+\big[\dot{\underline \mu}+\frac{1}{2}\,\underline \mu_{\,k,k}\big]\chi\\ \nonumber
&+& \big[-\frac{1}{2}\,\dot{\underline \mu}_k\big]\chi_k+\underline \mu \,\tau+\underline \mu^k\tau_k\,.
\end{eqnarray}
If now $\epsilon=\underline \mu$ and $\epsilon_k=\frac{1}{2}\,
\underline \mu_k$, then 
we find that
\begin{eqnarray}\label{B35}
\bar \delta h &=&\big\{h,G\big\}=\dot \epsilon+\epsilon_{,k}\\ \label{B36}
\bar \delta h_k &=&\big\{h_k,G\big\}=-\dot \epsilon_k-\epsilon_{,k}\\ \label{B37}
\bar \delta h_{ij} &=&\big\{h_{ij},G\big\}=(\dot \epsilon-\epsilon_{k,k})\delta_{ij}+\epsilon_{i,j}+\epsilon_{j,i}\,.
\end{eqnarray}

This is consistent with
\begin{equation}\label{B38}
\delta h^{\mu\nu}=\partial^\mu f^\nu+\partial^\nu f^\mu-\eta^{\mu\nu}\partial.f
\end{equation}
which is the form of the gauge transformation for eq. (\ref{B2}) discussed in ref. \cite{42}. Eq. (\ref{B38}) is in fact the 
linearized form of the diffeomorphism transformation. It remains to be seen if the linearized form of the gauge transformation 
of the full action of eq. (\ref{5}) implied by its first class constraints is given by eq. (\ref{B38}).

In the case $d=2$, the equation of motion for $G^\lambda_{\mu\nu}$ that follows from eq. (\ref{B1}) cannot be solved to express 
$G^\lambda_{\mu\nu}$ in terms of $h_{\mu\nu}$. However, if we were to set
\begin{equation}\label{B39}
G^\lambda_{\mu\nu}=V^\lambda\eta_{\mu\nu}+\bar G^\lambda_{\mu\nu}\,\,\,\,\,\,(\,\bar G^\lambda_{\mu\nu}\eta^{\mu\nu}=0\,)
\end{equation}
then eq. (\ref{B1}) when $d=2$ becomes
\begin{equation}\label{B40}
\tilde S_2=\int d^2x\big[-h^{\mu\nu}_{\,\,,\lambda}\,\eta_{\mu\nu} V^\lambda-h^{\mu\nu}_
{\,\,,\lambda}\bar G^\lambda_{\mu\nu}+(\bar G^\lambda_{\lambda\mu}\,\bar G^\sigma_{\sigma\nu}-
\bar G^\lambda_{\sigma\mu}\,\bar G^\sigma_{\lambda\nu})\,\eta^{\mu\nu}\big]
\end{equation}
and it is possible to express $\bar G^\lambda_{\mu\nu}$, the traceless part of $G^\lambda_{\mu\nu}$, 
in terms of $h_{\mu\nu}$. If we take
\begin{equation}\label{B41}
\frac{\delta \bar G^\lambda_{\mu\nu}}{\delta \bar G^\sigma_{\gamma \delta}}=\delta^\lambda_\sigma
\big[\,\frac{1}{2}(\delta^\gamma_\mu\,\delta^\delta_\nu+\delta^\gamma_\nu\,\delta^\delta_\mu)-\frac{1}{2}\eta^{\gamma \delta}\eta_{\mu\nu}\big]
\end{equation}
then the equation of motion for $\bar G^\lambda_{\mu\nu}$ results in
\begin{equation}\label{B42}
\bar G^\lambda_{\mu\nu}=-\frac{1}{2}(h^\lambda_{\mu,\nu}+h^\lambda_{\nu,\mu})+\frac{1}{4}\,(h^\rho_{\rho,\mu}\delta^\lambda_\nu+
h^\rho_{\rho,\nu}\delta^\lambda_\mu)+\frac{1}{2}\,h_{\mu\nu,}^{\,\,\,\,\,\,\lambda}+
\frac{1}{2}\eta_{\mu\nu}(h^{\lambda \rho}_{\,\,\,\,,\rho}-h^{\rho\,\,\lambda}_{\,\,\rho,})
\end{equation}
If eq. (\ref{B42}) is substituted back into eq. (\ref{B40}), then the action $\tilde S_2$ collapses down to
\begin{equation}\label{B43}
S_2=-\int d^2 x\, h^{\mu\nu}_{\,\,\,,\lambda}\eta_{\mu\nu}V^\lambda\,\,,
\end{equation}
showing the triviality of the spin-two field in two dimensions.

If we were to define 
$$\label{B44-B51}
h=h^{00}\,\,\,\,,\,\,\,\,h^1=h^{01}\,\,\,\,,\,\,\,\,\pi=-G^0_{00}\,\,\,\,,\,\,\,\,
\pi_1=-G^0_{01}\eqno (A37-A40)
$$
$$
\pi_{11}=-G^0_{11}\,\,\,\,,\,\,\,\,\xi=G^1_{00}\,\,\,\,,\,\,\,\,\xi_1=2G^1_{01}\,\,\,\,,\,\,\,\,
\xi_{11}=G^1_{11} \eqno (A41-A44)
$$
\addtocounter{equation}{8}
then eq. (\ref{B1}) when $d=2$ becomes
\begin{equation}\label{B52}
\tilde S_2=\int d^2 x \big[h_{,0}\pi+h^1_{,0}\pi_1+h^{11}_{,0}\pi_{11}-\xi(h_{,1}+\pi_1)-\xi_1(h^1_{,1}-\pi-\pi_{11})-
\xi_{11}(h^{11}_{,1}+\pi_1)\big]\,.
\end{equation}
These secondary constraints are
\begin{eqnarray}\label{B53}
\phi_1&=&h_{,1}+\pi_1\,,\\ \label{B54}
\phi&=&h^1_{,1}-\pi-\pi_{11}\,,\\ \label{B55}
\phi^1&=&h^{11}_{\,\,,1}+\pi_1\,.
\end{eqnarray}
These are analogous to the secondary constraints that arise from the first order EH action in two dimensions.
The PB of any two of these constraints vanishes.
\newpage
\section{Inclusion of Scalars}
We can supplement $S_d$ of eq. (\ref{1}) with
\begin{equation}\label{D1}
S_\phi= \frac{1}{2}\,\int d^dx\,\sqrt{-g}\, g^{\mu\nu}\,\partial_\mu \phi \,\partial_\nu \phi\,.
\end{equation}
The primary and secondary constraints of sections (IIB) and (IIC) are not altered by the inclusion of $S_\phi$. However, as a result of this
extra contribution to the action, the field $\phi$ has an associated momentum
\begin{equation}\label{D2}
p=h\,(\partial_0 \phi)+h^i\,(\partial_i \phi)
\end{equation}
which leads to a Hamiltonian density 
\begin{eqnarray}\label{D3}
\bar H_\phi &=& \left[\frac{p^2}{2h}+\frac{H^{ij}\partial_i\phi\,\partial_j \phi}{2}\right]-\frac{p\,h^i\partial_i\phi}{h}\\ \label{D4}
 & \equiv & H_\phi-\frac{p\,h^i\partial_i\phi}{h}\,.
\end{eqnarray}
Since
\begin{eqnarray}\label{D5}
\big\{\chi,\bar H_\phi \big\}&=& \bar H_\phi\\ \label{D6}
\big\{\chi_i,\bar H_\phi \big\}&=&-p\,\partial_i \phi
\end{eqnarray}
where $\chi$ and $\chi_i$ are the secondary constraints of eqs. (\ref{22},\ref{23}), the tertiary constraints of eqs. (\ref{51},\ref{52}) become
\begin{eqnarray}\label{D7}
T_i &=& \tau_i-p\,\partial_i \phi\\ \label{D8}
\bar T &=& \bar \tau+\bar H_\phi\,.
\end{eqnarray}
If we now set
\begin{equation}\label{D9}
T=\tau+H_\phi
\end{equation}
then
\begin{equation}\label{D10}
\bar T-T-\frac{h^i}{h}T_i=\bar \tau-\tau-\frac{h^i}{h}\tau_i\,.
\end{equation}
We now find that 
\begin{eqnarray}\label{D11}
f\big\{-p\,\partial_i\phi,-p\,\partial_j\phi\big\}g &=& g\partial_jf(-p\,\partial_i\phi)-f\partial_ig(-p\,\partial_j \phi)\\
f\big\{H_\phi,H_\phi\big\}g&=&(g\,\partial_if-f\,\partial_ig)\frac{H^{ij}}{h}(-p\,\partial_j \phi)\\
\big\{\chi,H_\phi\big\} &=& H_\phi\\
\big\{\chi_i,H_\phi\big\} &=& 0\\
\big\{\tau,H_\phi\big\} &=& \frac{1}{h} (H^{mk}H^{nl}\omega_{kl}-H^{mn}H^{kl}\omega_{kl})(\partial_m \phi \partial_n \phi)\\
\big\{\tau_i,-p\,\partial_j \phi\big\} &=& 0\\
\big\{\tau,-p\,\partial_j\phi\big\}&=&0
\end{eqnarray}
and
\begin{eqnarray}\label{D18}
&f&\big\{\tau_i-p\,\partial_i \phi,H_\phi \big\}\,\,g \\ \nonumber
&=& (fh)_{,i}\,\frac{H^{mn}}{2h}\,\phi_{,m}\phi_{,n}\,g+\frac{1}{2} (fH^{mn})_{\,,i}\, \phi_{,m}\phi_{,n}\,g+(pf)_{,i}\frac{p}{h}g
-\phi_{,i}\,(fH^{mn} \phi_{,m} \,g)_{,n}\\ \nonumber
&=&\left[\frac{1}{h}(fh)_{,i}\,g-fg_{,i}\right] H_{\phi}+\left(\frac{fg\,p^2}{2h}\right)_{,i}-\left(fgH^{mn}\phi_{,i}\phi_{,m}\right)_{,n}+\frac{1}{2}(fgH^{mn}\phi_{,m}
\phi_{,n})_{,i}
\end{eqnarray}
The total divergences appearing in eq. (\ref{D18}) can be neglected. It now follows from eqs. (\ref{D11}-\ref{D18}) that the PBs of eqs. (\ref{56}-\ref{62},\ref{66},\ref{67}) (which arise when dealing with pure gravity defined by eq. (\ref{1})) can be modified to accommodate the scalar field by simply replacing $\tau$ and $\tau_i$ by $T$ and $T_i$ respectively.
This result shows that the gauge transformation implied by the first class constraints in pure gravity and in pure gravity supplemented by a free scalar field are clearly related.
\newpage
\end{document}